\documentclass[prb,twocolumn,showpacs,preprintnumbers,amsmath,amssymb,floatfix]{revtex4}

\usepackage{graphicx}
\usepackage{dcolumn}
\usepackage{bm}
\usepackage{amsmath} \usepackage{amsfonts} \usepackage{amssymb}

\begin{document}
\title{Anomalous thickness dependence of the Hall effect in
       ultrathin Pb layers on Si(111)}
\author{I. Vilfan$^{1,2}$}
\email{igor.vilfan@ijs.si}
\author{M. Henzler$^1$}
\email{henzler@fkp.uni-hannover.de}
\author{O. Pfennigstorf$^1$}
\author{ H. Pfn\"{u}r$^1$}
\email{pfnuer@fkp.uni-hannover.de}
\affiliation{$^1$Institut f\"{u}r Festk\"{o}rperphysik, Universit\"{a}t Hannover,
 Appelstrasse 2, D-30167 Hannover, Germany}
\affiliation{$^2$J. Stefan Institute, Jamova 39, SI-1001 Ljubljana, Slovenia}
\date{\today}

\begin{abstract}
The magnetoconductive properties of ultrathin Pb films deposited on
Si(111) are measured and compared with density-functional 
electronic band-structure calculations on two-dimensional,
free-standing, 1 to 8 monolayers thick Pb(111) slabs.
A description with free-standing slabs is possible because it turned out that  
the Hall coefficient is independent of the substrate and of
the crystalline order in the film.
We show that the oscillations in sign of
the Hall coefficient observed as a function of film thickness can be
explained directly from the thickness dependent variations of the
electronic bandstructure at the Fermi energy.  
\end{abstract}

\pacs{
73.50.Jt  % Electronic transport phenomena in thin films:
          % Galvanomagnetic and other magnetotransport effects
73.61.At  % Electrical properties of specific thin films: 
          % Metals and metallic alloys
73.20.At  % Electron states at surfaces and interfaces
71.15.Mb  %Density functional theory
}
\maketitle
%==========================================================
%                                                         %
%               \section{Introduction}                    %
%                                                         %
%==========================================================

The electronic properties of crystalline bulk materials are well
understood. 
However, as soon as the lateral dimension in any direction is reduced
to a few atomic distances, each  electron energy band splits into a 
set of discrete subbands and the spatial confinement leads to quantum-size effects (QSE).
The quantum-size confinement affects in particular the electron transport properties.
In an early theoretical approach Trivedi and Ashcroft \cite{TA88}
considered transport of electrons in a metal film, confined in a two-dimensional
potential well and scattered by impurities in the film and/or by surface roughness.
They predicted a saw-tooth like variation of the conductivity with the
film thickness with a period of half the Fermi wavelength.
Such oscillations have not been observed in the experiment so far.
Another source of QSE-induced conductivity oscillations 
%with increasing thickness of ultrathin films 
is the layer-by-layer growth of, e.g., Pb films on Si(111)$(7\times 7)$
or Si(111)-Au$(6\times 6)$ when the in-plane conductivity shows
oscillations with the period of the interlayer spacing, associated with
the degree of roughness of the surface.\cite{JBKL,PH01} 
An even more spectacular, but so far unexplained, observation was the reversal of sign 
in the Hall coefficient $R_H$ of epitaxial Pb films on Si(111), grown on a surface 
that was pre-covered with about one monolayer (ML) of Ag which shows a 
$(\sqrt{3}\times \sqrt{3})$ reconstruction.\cite{JHB96} 

In this report, we address the general issue of charge transport and
Hall effect in ultrathin metallic layers both theoretically and experimentally.
We focus on Pb films deposited on Si(111), although the conclusions
should be generally valid for ultrathin metallic films in the region of QSE. 
In order to clarify the underlying mechanisms, we combined in our study theoretical 
investigations by density-functional
calculations with measurements of conductivity and of magnetoconduction of ultrathin
Pb films on a $(7\times 7)$ reconstructed Si(111) substrate.

%==========================================================
%                                                         %
%              \section{Experimental Setup}               %
%                                                         %
%==========================================================

Pb films with a thickness between 1 and 12 monolayers (ML) have been 
deposited in UHV on a well annealed 
Si(111)$-(7\times 7)$ substrate at low temperatures (15 to 25 K).
The  thickness, crystalline order and defects have been monitored with 
high-resolution low-energy electron diffraction (SPA-LEED).\cite{Pet01}
4 Mo contacts have been pre-deposited for the measurements.
After the pseudomorphic first Pb layer, up to 4 Pb layers were amorphous,
whereas thicker layers always turned out to be crystalline. 
By varying temperature and thus the growth conditions, the influence of order on the 
conductive and magnetoconductive properties was tested.  
For details on film preparation and ordering in the films, see Ref.~\onlinecite{Pfen02}.

Before starting the Hall effect measurement the films have been annealed 
close to temperature where the decrease of conductance indicated a 
break-up of the film into non-contiguous islands.
Even after annealing, films with a thickness $d < 4$ML were strongly disordered
whereas thicker films grew epitaxially with the lattice constant of bulk Pb. 
While the conductance measurements, which provide information on the charge-carrier 
scattering, were to some extent sensitive to the degree of order, 
the  oscillatory behavior of the Hall coefficient turned out to be remarkably 
insensitive. 
Therefore, it seems to be justified as a first approach to compare the experimental 
results of the Hall coefficient with those obtained theoretically for free-standing 
Pb slabs. 
Indeed, as we will show below, such calculations are able 
to identify the basic origin of the oscillations. 

Conductance and magnetoconductance measurements were performed 
in a DC magnetic field up to 4 T at various temperatures down to about 7 K. 
In the van-der-Pauw arrangement for conductance \cite{vdP58} it has been checked
that the magnetoconductance was completely symmetric.
In the contact arrangement for Hall effect there was a sizable contribution 
from magnetoconductance due to deviations from a perfect quadratic arrangements 
of the contacts.
Therefore the asymmetry was taken as the contribution of the Hall effect
to the measured voltage. The Hall coefficient has been derived
from this asymmetry at $\pm 4T$.

%==========================================================
%                                                         %
%            \section{Method of Calculation}              %
%                                                         %
%==========================================================

In the model calculations we treat the Pb layers as  free-standing
slabs and neglect, as mentioned, the effect of the substrate. 
Within the model we calculate, as a function of layer thickness, the densities of 
states at the Fermi energies, Fermi velocities, and effective-mass tensors. 
From these quantities, together with the experimental data for the conductivity, 
$\sigma_0^{\rm exp}$, the elastic  scattering times are calculated and the
Hall coefficient, $R_H$ is derived. 

The Hall coefficient in the low-field limit is
\begin{equation}
    R_H = \frac{\sigma_H}{\sigma_0^2}
\end{equation}
where $\sigma_0$ and $\sigma_H$ are the electrical and Hall
conductivities of a two-dimensional slab.
Adapting the three-dimensional expressions \cite{SAT92} to
two dimensions we have
\begin{equation}
    \sigma_0 = \frac{e^2}{V\hbar^2} \sum_{n,k} \tau_n(k) \left[\nabla_k\epsilon_n(k)
    \right]^2 \left(-\frac{\partial f(\epsilon)}{\partial \epsilon}\right)
   \label{el_cond}
\end{equation}
and
\begin{eqnarray}
  \sigma_H = \frac{ e^3}{V}\sum_{n,k} \tau_n(k)^2 \vec{v}_n(k)[{\bf Tr}({\bf M}^{-1})
  - {\bf M}^{-1}] \vec{v}_n(k) \nonumber \\
  \times\left(-\frac{\partial f(\epsilon)}{\partial \epsilon}\right).
  \label{hall_cond}
\end{eqnarray}
${\bf M}^{-1}$ is the inverse mass tensor with the elements
\begin{equation}
   ({\bf M}^{-1})_{i,j} = \frac{1}{\hbar^2} \frac{\partial^2 \epsilon_n(k)}
   {\partial k_i \partial k_j},
\end{equation}
$V$ the slab volume, $\vec{v}_n(k) = \frac{1}{\hbar} \vec{\nabla}\epsilon_n(k)$ 
the electron group velocity,  $\epsilon_n(k)$ the energy of an 
electron with momentum $k$ in the subband $n$, and $f$ the Fermi function.
The sums in Eqs.~(\ref{el_cond}) and (\ref{hall_cond}) run over the
wavevectors $k$ in the \textit{two-dimensional} hexagonal Brillouin zone
and over the electron subband index $n$. 
(Each atomic electron level splits into $d$ two-dimensional electron
subbands where $d$ is the number of monolayers in the slab.)
The spin degeneracy is included in the prefactors. 
As will be seen below, the relaxation time $\tau_n(k)$ of thin slabs at low temperatures 
is dominated by elastic scattering on the thickness fluctuations of the Pb film 
(roughness of the interface to the substrate).  
Therefore, we will take the limit of short-wavelength surface height fluctuations 
(with their correlation length $\xi \ll 1/k_F$, $k_F$ being the Fermi wavevector). 
In this case  the relaxation time is $k-$independent.\cite{C90}
Since no detailed information on the scattering mechanisms and on the interface structure
is available, we assume that $\tau$ is also independent of the subband index $n$.
Then, $R_H$ is independent of $\tau$ and is determined solely by the electron
bandstructure at the Fermi energy $\epsilon_F$.
\begin{figure}[tb]
  \includegraphics{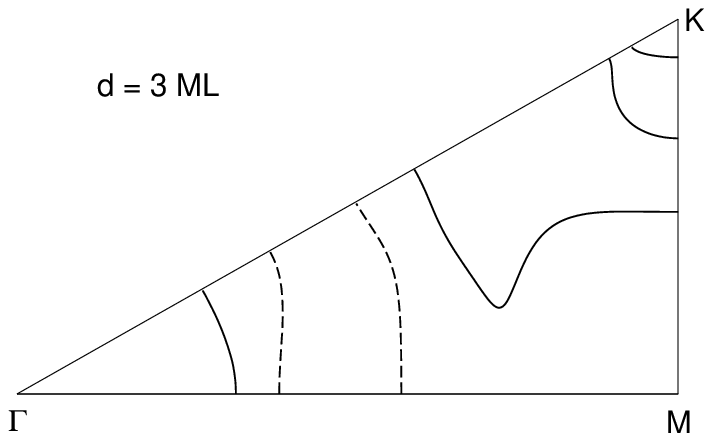}\\
  \includegraphics{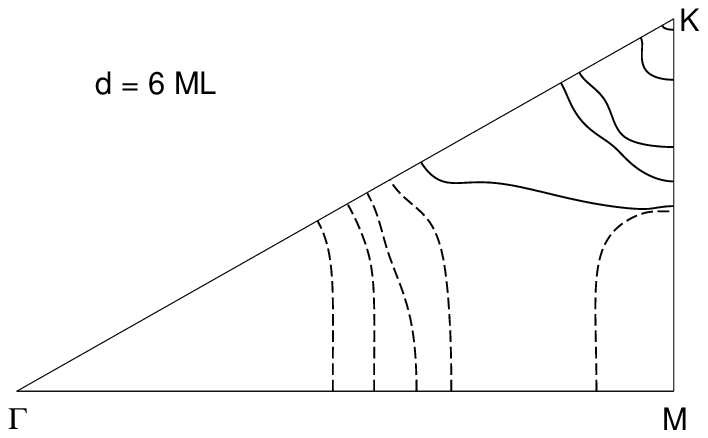}
  \caption{\label{fermi}
  Fermi lines of relaxed 3 and 6 ML thick free-standing Pb slabs.
  The solid lines encircle electron and dashed lines hole pockets.
  }
\end{figure}

The electronic bandstructure calculations were performed on 
relaxed 1 to 8 monolayers thick (111) oriented Pb 
slabs with the in-plane lattice constant $a = 3.50$ {\AA},
separated by $\approx 10$ {\AA} of vacuum.
The electron band energies and the total energy have been calculated using
the full-potential linearized augmented plane-wave method within the  WIEN2k 
code \cite{BSL97} in the local-density approximation.\cite{PW92} 
A Pb muffin-tin radius of 2.6 a.u. and a tetrahedral mesh of 623 $k-$points in the
irreducible part of the full Brillouin zone were used in the self-consistent electronic
structure calculations.
The kinetic-energy  cutoff was set to $E_{\rm max}^{\rm wf} = 7.3$ Ry
and the plane-wave expansion cutoff to  $E_{\rm max}^{\rm pw} = 100$ Ry. 
Later, in the magnetoconduction calculations, we used about 4000 $k$ points
in the irreducible part of the two-dimensional Brillouin zone.
 
Several electron subbands cross the Fermi energy and there are several
\textit{Fermi lines} in the two-dimensional Brillouin zone. 
As typical examples, Fig.~\ref{fermi} shows the Fermi lines of 3 and
6 ML thick slabs. 
The number of Fermi lines is an increasing function of the slab thickness. 
Only electrons close to the Fermi lines take part in the transport and  
$\sigma_0/\tau$ and $\sigma_H/\tau^2$ are obtained after numerical
integration of expressions (\ref{el_cond}) and (\ref{hall_cond}) along
the Fermi lines.
Whereas $\sigma_0$ is always proportional to the density of states at
the Fermi lines , $\sigma_H$ of a subband is negative if the Fermi
line makes a closed loop around occupied states (electrons) and positive if the
loop encircles empty states (holes) in a subband.  
It is the balance between the number of electron and hole states at $\epsilon_F$ 
together with their effective masses which determines the sign of $R_H$.  
As a consequence, $R_H$ is sensitive to the details of the bandstructure.
%==========================================================
%                                                         %
%                    \section{Results}                    %
%                                                         %
%==========================================================

\begin{figure}[tb]
 \includegraphics{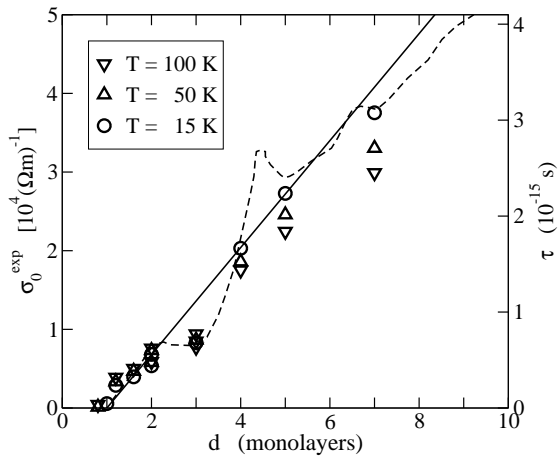}
  \caption{\label{sigma}
  Experimental conductivity $\sigma_0$ and the calculated average relaxation 
  time $\tau$. 
  The symbols are for annealed Pb films on Si(111)$-(7\times7)$, measured
  at three different temperatures,  the dashed line is for epitaxial Pb on
  Si(111)-Ag$(\sqrt{3}\times\sqrt{3})$ at $T=20$ K from Ref.~\cite{JHB96}
  and the solid line is a linear fit to the low-temperature data for $d<6$ML.
  The main process limiting the conductivity and the relaxation
  time in Pb ultrathin films is scattering on the interface fluctuations,
  the disorder within the film plays only a minor role.
  }
\end{figure}

Fig.~\ref{sigma} shows the experimental conductivities $\sigma_0^{\rm exp}$
of annealed Pb films on Si(111)$-(7\times 7)$ together with the data of 
epitaxial Pb on Si(111) covered with Ag.\cite{JHB96}
Three properties of these data indicate that interface and/or surface scattering is 
the dominant mechanism of (elastic) electron scattering in this system. 
Firstly, only a  weak temperature dependence of $\sigma_0^{\rm exp}$ is found 
below 100 K. Secondly, there is only a weak dependence of conductivity on crystallinity 
and order in the films. Annealing of amorphous films grown at 15 K resulted in an 
increase of conductivity, but this increase never exceeded a factor of 2.5. 
Thirdly, the roughly linear thickness dependence of $\sigma_0^{\rm exp}$ is again 
consistent with the assumption just made that the dominant scattering mechanism
in these ultra-thin films is diffuse scattering at interface fluctuations.\cite{C90,CL99} 
For thicker samples the scattering at imperfections within the film becomes noticeable, 
and the conductivity starts to deviate from the linear behavior and eventually saturates
in the bulk value, where the scattering on lattice defects and at higher temperatures 
also on phonons limit the conductivity.

We measured $\sigma_0^{\rm exp}$ for arbitrary film thickness $d$.
At non-integer thicknesses the film surface is covered with islands
and their borders (i.e., steps) act as scattering centers for the electrons.
Therefore one would expect a dip in $\sigma_0^{\rm exp}$ for
non-integer thicknesses. However, a small effect of this kind has only been observed 
at $d>12$ ML. This suggests that the typical island sizes exceed the charge-carrier 
mean free path for very thin layers and electron scattering at the vacuum interface 
is not important for films with $d<10$.

Clearly seen is an offset in the conductivity towards $d = 1$ ML. 
The onset of conductivity due to percolation of the evaporated Pb at a
surface temperature of 15 K was found to be at 0.8 ML. 
The very low conductivity of this first monolayer 
indicates that the main transport mechanism in this layer is hopping between
localized states. 
Indeed, the conductivity shows a temperature-activated behavior for
$d=1$ ML.\cite{Pfen02} 
The first Pb monolayer on Si(111)$-(7\times 7)$ has the periodicity 
of the substrate  \cite{Pet01} and may incorporate the adatoms 
of the $(7\times 7)$ substrate. This might be the reason for the very low
conductivity and for the localization of charge carriers.

The close agreement of the conductivity data obtained in our measurements with those of 
Ref.~\onlinecite{JHB96} is not at all self-evident, since our measurements have been 
carried out on Pb films that are expected to be much more disordered than those grown 
epitaxially on Si(111)-Ag$(\sqrt{3}\times\sqrt{3})$. 
Since the thickness dependence is almost the same for Pb films prepared on both 
substrates, we must conclude that interface scattering plays the same dominant role on 
both substrates. 
This means on the other hand that the degree of order in these ultrathin films is of 
secondary importance for electronic transport. We also note that the Ag interface layer 
obviously does not contribute to the measured conductance.

In contrast to the experimentally measured $\sigma_0^{\rm exp}$, the calculations of 
$\sigma_0/\tau$ and $\sigma_H/\tau^2$ were performed only for integer $d$.
Surprisingly, $\sigma_0/\tau$ turned out to be almost independent of
the slab thickness in the investigated range between 2 and 8 ML, 
\begin{equation}
   \frac{\sigma_0}{\tau} \approx 1.1 \times 10^{21} {\rm(\Omega m s)}^{-1}.
   \label{sigcalc}
\end{equation}

With increasing number of monolayers the number of subbands increases 
and thus the conductance but not the ratio $\sigma_0/\tau$. 
This means that the contributions of newly emerging subbands to $\sigma_0/\tau$ as a 
function of layer thickness are very similar. 
Eq.~(\ref{sigcalc}) implies that $\tau$ is proportional to $\sigma_0$.
Using the experimental values for $\sigma_0$, we determined $\tau$,
also shown in Fig.~\ref{sigma}. 

It is interesting to look at the charge-carrier mean free path 
$\lambda = v_F \tau$. 
From the bandstructure we get the Fermi velocity  $v_F \sim 1.2 \times 10^6$ m/s and, 
using the values for $\tau$ from Fig.~\ref{sigma}, we see that $\lambda$ is of the order 
1 to 2 film thicknesses.
It is therefore tempting to say that the electrons scatter on the 
film surfaces. However, the charge carriers
\textit{scatter only on imperfections} (steps, misplaced atoms etc.) on the surfaces
and not on a perfectly flat surface.
For $d=1$ ML the mean free path is shorter than the lattice spacing and
the wave-vector description of the electron states breaks down.

The relaxation time causes broadening of the electron energy levels in the vicinity of 
the Fermi energy.
Therefore, $R_H$ was calculated as $\langle\sigma_0/\tau\rangle /
\langle\sigma_H/\tau^2\rangle$ where $\langle\rangle$ is an average over the Gaussian 
distribution  with the half-width $\delta \sim \hbar/\tau$.

\begin{figure}[tb]
  \includegraphics{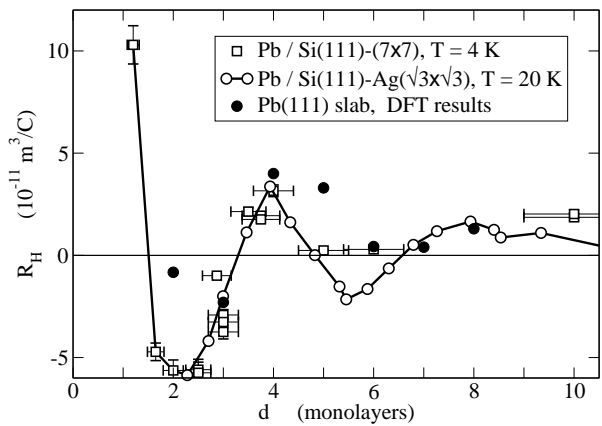}
  \caption{\label{hall}
  Quantum-size induced 
  oscillations of the Hall coefficient with the thickness $d$.
  Solid circles: calculations for a free-standing Pb slab;
  open symbols: experimental values for Pb films on two
  different Si(111) substrates (open circles are from Ref.~\onlinecite{JHB96})
  Unless explicitly indicated, the experimental error bars are 
  of the size of the symbols. The estimated error bars of
  calculated $R_H$ are $\pm 1\times 10^{-11}$ m$^3$/C and 
  come mainly from uncertainty in the broadening of the
  energy levels due to scattering. The solid line is a guide for
  the eye through the experimental points.   
  } 
\end{figure}

The experimental results (open symbols)  
of the Hall coefficient are compared in Fig.~\ref{hall} with those calculated (solid circles).
Whereas the calculations have been done for perfectly periodic slabs without substrate, 
the measurements refer to epitaxial films with different degrees of disorder. 
Clearly seen are the anomalous oscillations of $R_H$ as a function of film
thickness $d$ for both sets of experimental data.
Our experimental results (squares) agree very well within the experimental error with 
the results of Ref.~\onlinecite{JHB96} (open circles in Fig.~\ref{hall}). 
After having found close agreement for the data of electrical conductance, the agreement 
of the results for $R_H$ are not surprising, since the disorder within the film
% expected (see Eqs. (1), (2) and (3)) to 
plays a negligible role. On the other hand, this result justifies our assumption of 
scattering times independent of $k$ and $n$. 
%==========================================================
%                                                         %
%                  \section{Discussion}                   %
%                                                         %
%==========================================================

$R_H$, calculated in this paper without any adjustable parameter, 
reproduces the oscillating behavior seen in the experiment semi-quantitatively and 
demonstrates that the main reason for this behavior are the thickness dependent changes 
in the band structure. 
The deviations at $d=2$ and $d=5$ may be a consequence of the approximation of equal 
relaxation times for all subbands. Especially in the case $d=2$ also the $7\times 7$ 
periodicity of the substrate, which was neglected in the slab calculation, might play a role.
Our calculations also show that  $\sigma_0/\tau$ is a slowly varying 
function of energy. Therefore it is not very sensitive to the exact location of the Fermi 
energy within the band structure, and the conductance increases smoothly with film 
thickness, and with only small variations during deposition.
In contrast, the calculated $\sigma_H/\tau^2$ is a strongly energy-dependent
function and most of the energy dependence of $R_H$ comes from  
$\sigma_H/\tau^2$, therefore $R_H$ is sensitive to averaging. 
%==========================================================
%                                                         %
%                   \section{Conclusions}                 %
%                                                         %
%==========================================================

To conclude, we have shown that only a quantitative bandstructure 
analysis, done here with DFT, allows an understanding 
of the anomalous magnetoconductive properties of ultrathin Pb films.
Oscillations in $R_H$ as a function of film thickness 
are the consequence of competing -- and to a large extent compensating --
contributions of electrons from different subbands and
cannot be explained with any model which does not take into
account the details of the electronic bandstructure. 
The relaxation time and thus the interface fluctuations play a subordinate 
role, they affect $R_H$ through the width of the distribution function
and through the subband-dependence of the relaxation time, which was 
neglected in this paper.
Although our calculations with free-standing films already show all 
qualitative features observed in experiment, it is obvious that this method could 
be further refined by explicitly taking into account the underlying substrate. 
This might help to understand the specific role of the
metal-semiconductor interface considered here. 
Generally, this interplay between calculations and experiments is of
primary importance since inner interfaces cannot be varied easily in experiments. 
In this context, also the open problem of the very low conductivity and the high
$R_H$ of the first monolayer may be solvable. 
This involves most likely hopping conduction between localized states in the Pb 
film on reconstructed Si(111) substrate.

%\acknowledgments
The authors are grateful to Prof.~Matthias Scheffler for interesting 
discussions and Prof.~Peter Saalfrank for sending us his results prior
to publication. I.V.~would like to acknowledge the hospitality of the 
Institut f\"{u}r Festk\"{o}rperphysik in Hannover.
This work was supported in part by the Deutsche Forschungsgemeinschaft.
%==========================================================
%                                                         %
%                     R E F E R E N C E S                 %
%                                                         %
%==========================================================

\end{document}